\documentclass[11pt]{article}

\newcommand{\be}{\begin{equation}}
\newcommand{\ee}{\end{equation}}
\newcommand{\ba}{\begin{eqnarray}}
\newcommand{\ea}{\end{eqnarray}}
\newcommand{\nn}{\nonumber}
\newcommand{\al}{\alpha}

\newcommand{\G}{\Gamma}

\newcommand{\MSsch}{{\overline{\rm MS}}}
\begin{document}
\phantom{}
\vspace{1.1cm}
\thispagestyle{empty}
\begin{flushright}
MZ-TH/01-29\\
October 2001
\end{flushright}
\vspace{0.5cm}
\begin{center}
{\Large\bf Heavy quark production near the threshold in QCD}\\[1truecm]
{\large \bf A.A.Pivovarov}\\[1truecm]

Institut f\"ur Physik, Johannes-Gutenberg-Universit\"at,\\
Staudinger Weg 7, D-55099 Mainz, Germany\\
and \\
Institute for Nuclear Research of the\\
Russian Academy of Sciences, Moscow 117312, Russia
\end{center}

\vspace{1truecm}
\begin{abstract}
\noindent Theoretical results for the cross section
of heavy quark production near the threshold at NNLO of NRQCD
are briefly overviewed.
\end{abstract}

\vspace{6truecm}
\begin{center}
{\it Talk given at 10th Lomonosov conference on Elementary Particle Physics,\\
August 2001, Moscow, Russia}
\end{center}

\newpage
\section{Introduction}
The heavy quark production near thresholds will be thoroughly 
investigated at future accelerators, e.g.~\cite{exp}. 
One can study the $q\bar q$ systems near 
threshold in $e^+e^-$ annihilation~\cite{ttee,ttee1}
and $\gamma\gamma$ collisions~\cite{ttgg,ttgg1}. In 
$e^+e^-$ annihilation the production vertex is 
local (the electromagnetic current in case of photon
and/or neutral weak current in case of $Z$-boson),
the basic observable is a production cross section
which is saturated by S-wave (for the vector current).
In $\gamma\gamma$ collisions the production vertex is 
nonlocal and both S- and P-waves can be studied for different helicity 
photons, the number of observables is larger 
(cross sections $\sigma_S$, $\sigma_P$, S-P interference).
In $e^+e^-$ annihilation high precision data are already 
available for $b\bar b$ and expected for $t\bar t$. 
The $\gamma\gamma \to q\bar q$ experiments are planned.

\section{Features of theoretical description}
Heavy quarks near the production threshold moves slowly 
that justifies the use of the nonrelativistic quantum mechanics 
for their 
description~[6-8].
Being much simpler than the comprehensive relativistic treatment 
of the bound state problem
with Bethe-Salpeter amplitude~\cite{BetheSal},
the nonrelativistic approach allows one to take into account 
exactly such an essential feature
of the near-threshold dynamics as Coulomb interaction~\cite{VL}.
For unstable heavy quarks with a large decay width it is possible to
compute the cross section near threshold point-wise in energy
because the large decay width suppresses the long distance
effects of strong interaction \cite{FK}.
The $t\bar t$-pair near the production 
threshold is just a system that satisfies the requirement of being
nonrelativistic.
Therefore the description of $t\bar t$-system near the production
threshold $\sqrt{s}\approx 2m_t$ ($\sqrt{s}$ is a total energy
of the pair,
$m_t$ is the top quark mass) is quite precise within 
nonrelativistic QCD (NRQCD).
Reasons for this accuracy are related to the large mass of the top: 
\begin{itemize}
\item
The top quark is very heavy $m_t=175~{\rm GeV}$ \cite{PDG}
and there is an energy region of about $8-10$ GeV near the 
threshold where the nonrelativistic approximation
for the kinematics is very precise.
For $\sqrt{s}=2m_t+E$ with $|E|< 5~{\rm GeV}$ the quark velocity is small 
\be
v=\sqrt{1-\frac{4m_t^2}{s}}=\sqrt{1-\frac{4m_t^2}{(2m_t+E)^2}}\simeq
\sqrt{\frac{E}{m_t}} < 0.15 \ll 1 \, .
\ee
Relativistic effects are small and can be taken into account
perturbatively in $v$ (even in $v^2$).
\item
The strong coupling constant at the high energy scale is small 
$\al_s(m_t)\approx 0.1$ that makes 
the perturbative mapping of QCD onto the low energy 
effective theory (NRQCD) numerically precise.
\item
The decay width of top quark is large, $\G_t=1.43$ GeV;
infrared (small momenta) region is suppressed and PT calculation 
for the cross section near the threshold is reliable 
point-wise in energy.
\end{itemize}
Because $\al_s \sim v$ and the ratio $\al_s/v$ is not small,
the Coulomb interaction is enhanced.
The ordinary perturbation theory for 
the cross section (with free quarks as the lowest order
approximation)
breaks down and all terms of the order $(\al_s/v)^n$ should be summed.
The expansion for a generic observable $f(E)$ in this kinematical region
has the form
\be
\label{generic}
f(E)=f_0(\al_s/v)+\al_s f_1(\al_s/v)+\al_s^2 f_2(\al_s/v)+\ldots
\ee
where 
$f_i(\al_s/v)$ are some (not polynomial) functions of the ratio
$\al_s/v$, $f_0(\al_s/v)$ is a result of the pure Coulomb approximation
(or a kind of its improvement).
The expansion in $\al_s$ in eq.~(\ref{generic})
takes into account the perturbative QCD corrections to the parameters 
of NRQCD and relativistic corrections (in the regime $v\sim \al_s$).

For the $e^+e^-\to t\bar t$ process mediated by the photon 
the NNLO analysis is well known.
The basic quantity is the vacuum polarization
function
\be
\Pi(E)=i \int \langle T j_{em}(x)j_{em}(0)\rangle e^{iqx} dx, 
\quad q^2=(2m_t+E)^2\, .
\ee
Near the threshold (for small energy $E$) NRQCD is used.
The cross section is saturated with 
S-wave scattering. In this approximation 
the polarization function near the threshold to the NNLO
accuracy in NRQCD is given by 
\be
\Pi(E)=\frac{2\pi}{m_t^2}C_h(\al_s)C_{\cal O}(E/m_t) G(E;0,0)\, .
\label{Rv}
\ee 
The pole mass definition is used for $m_t$ (e.g. \cite{Tar}),
$\al_s$ is the strong coupling constant.
$C_h(\al_s)$ is the high energy coefficient.
$G(E;0,0)$ is the nonrelativistic Green function (GF).
The quantity $C_{\cal O}(E/m_t)$ describes the contributions 
of higher dimension operators within the effective theory
approach. These contributions have, in general, a different
structure than the leading term. To the NNLO of NRQCD the contribution
of higher dimension operators  
can be written as a total factor $C_{\cal O}(E/m_t)$
for the leading order GF,
$C_{\cal O}(E/m_t)=1-4E/3m_t$.
The polarization function near the threshold (\ref{Rv})
contains expansions in small parameters
$\al_s$ and/or $v$, cf. eq.~(\ref{generic}).
The leading order approximation 
of the low energy part is
the exact Coulomb solution for the Green function.

The nonrelativistic Green's function $G(E)=(H-E)^{-1}$ is determined  
by the nonrelativistic Hamiltonian
\be
\label{hamNR}
H=\frac{p^2}{m_t}+V(r)
\ee
describing dynamics of the $t\bar t$-pair near the threshold.
The most complicated part of Hamiltonian (\ref{hamNR})
to find is the heavy quark static
potential $V_{pot}(r)$ entering 
into the potential $V(r)$. 
The static potential $V_{pot}(r)$ is computed in 
perturbation theory and can be written in the form
\be
V_{pot}(q)=-\frac{4 \al_{\rm v}(r)}{3 r}
\ee
that gives a definition of the effective charge $\al_{\rm v}$ related 
to the $\MSsch$-scheme coupling constant
\be
\label{a1a2}
\al_{\rm v}(\mu)=\al_s(\mu)(1+a_1 \al_s(\mu)+a_2 \al_s(\mu)^2)\, .
\ee
Coefficients $a_{1,2}$ are known~[14-15].
The effective coupling $\al_{\rm v}$ is nothing but a coupling 
constant in some special subtraction scheme.
The coefficient $a_2$ allows one to find the effective $\beta$-function 
$\beta_{\rm v}$ for the evolution of the coupling $\al_{\rm v}$ at NNLO.  

High energy coefficient $C_h(\al_s)$ is given by the expression
\be
\label{hic}
C_h(\al)=1-\frac{16}{3}\frac{\al_s}{\pi}+\left(\frac{\al_s}{\pi}\right)^2
\left(-\frac{140}{27}\pi^2\ln\frac{\mu_f}{m_t}+c_2\right)\, .
\ee
The NLO result for $C_h(\al_s)$ has been known since long 
ago~[17-18].
At NNLO there appears a term proportional to the logarithm
of the factorization parameter $\mu_f$
that separates long and short distances 
(or large and small momenta) within the effective theory approach.
The finite ($\mu_f$ independent) 
coefficient $c_2$ is known \cite{coef1,coef2}.
An explicit dependence of high and low energy quantities
on the factorization scale $\mu_f$ 
is a general feature of effective theories which are  
valid only for a given region of energy.
Physical quantities are factorization scale independent.
In NRQCD the $\mu_f$ dependence cancels between
Green's function and the high energy coefficient $C_h$.

The main dynamical quantity in description of the $t\bar t$ system 
near the threshold is the nonrelativistic Green's function
$G=(H-E)^{-1}$.
The Hamiltonian is represented in the form~[22-24]
\be
H=H_C+\Delta H, \quad H_C=\frac{p^2}{m_t}-\frac{4\al_s}{3r}
\ee
with $\Delta H$ describing high order corrections. Constructing 
the Green's function is straightforward and can be done
analytically within perturbation theory near Coulomb Green's 
function $G_C(E)$ or numerically for complex values of $E$
that can be used to describe the production of particles with nonzero 
width~[25-33].

The analytical solution for Green's function is perturbative 
in $\Delta H$ 
\be
G=G_C-G_C \Delta H G_C +G_C \Delta H G_C \Delta H G_C-\ldots
\ee
Results are presented basically as an expansion
in consecutive orders
\be
G=G_0+\Delta G_1 +\Delta G_2
\ee
to check the convergence of the approximations.
The leading order is given by Coulomb approximation, $G_0=G_C$.
At NLO the quantity $\Delta G_1$ takes into account the corrections from 
the static potential $V_{pot}(r)$ related to $a_1$ coefficient 
in eq.~(\ref{a1a2}).   
At the NNLO the quantity $\Delta G_2\sim O(\al_s^2)$ in a sense of 
eq.~(\ref{generic}) accounts for $\al_s^2$
terms in the static potential $V_{pot}(r)$ 
($a_2$ coefficient in eq.~(\ref{a1a2})) and relativistic $v^2$ corrections.
It also contains a second iteration of the $O(\al_s)$ term 
from $V_{pot}(r)$. The $\MSsch$-scheme for the static potential
$V_{pot}(r)$ is mainly used in the solution.
The numerical results for GF obtained by different authors agree 
with each other~\cite{revHoa}.

\section{Physical results}
For the $b\bar b$ system an accurate description of the spectrum in terms
of moments
\[
{\cal M}_n=\int_{4m_b^2}^\infty \frac{\rho(s)ds}{s^n}
\]
is obtained in the near-threshold Coulomb
PT calculations.
This analysis gave the best determination of the numerical value for 
the $b$ quark mass
\[
m_b=4.80\pm 0.06~{\rm GeV}\, .
\]
For the $t\bar t$ system 
the top quark width $\G_t$ plays a crucial role
in the calculation of the production
cross section near the threshold. At the formal level 
the width is taken into account by a shift of the energy variable
$E$. The mass operator of the top quark is 
approximated by the expression 
$M=m_t-i\G_t/2$. Then the kinematical variable $s-4m_t^2$
relevant to the near-threshold dynamics
is substituted with $s-4M^2$ ($\sqrt{s}=E+2m_t$)
and one finds 
\[
s-4M^2=4m_t(E+i\G_t)+E^2+\G_t^2\, .
\]
Neglecting higher orders in $E$ and $\G_t$ one obtains a recipe
for taking into account the width $\G_t$ by the shift $E\to E+i\G_t$.
The dispersion relation for the vacuum polarization function $\Pi(E)$
has the form 
\[
\Pi (E)=\int \frac{\rho(E')dE'}{E'-E}\, .
\]
With the shift recipe one finds 
\be
\label{dispr}
\sigma(E)\sim {\rm Im}~\Pi (E+i\G_t)
= {\rm Im}~\int \frac{\rho(E')dE'}{E'-E-i\G_t}
= \G_t\int \frac{\rho(E')dE'}{(E'-E)^2+\G_t^2}\, .
\ee
Because the point $E+i\G_t$ lies sufficiently far from the positive semiaxis
(and the origin) in the complex energy plane
the cross section eq.~(\ref{dispr}) is calculable point-wise in 
energy. The hadronic cross section $\sigma(E)$ was obtained 
by many authors~\cite{revHoa}.
The normalized cross section for typical values 
$m_t=175$~GeV, $\G_t=1.43$~GeV, $\al_s(M_Z) = 0.118$ has the  
characteristic points which are usually considered as basic 
observables. They are:
$E_p$ -- the position of the peak in the cross section
and $H_p$ -- its height. In the limit of the small $\G_t$ (at least for $\G_t$
that is smaller than the spacing between 
the first two Coulomb poles) one would have 
$E_p\sim E_0$ and $H_p\sim |\psi_0(0)|^2$.
The actual value of $\G_t=1.43$~GeV is larger than the 
spacing $|E_0-E_1|\sim m_t \al_s^2/3\approx 0.6~{\rm GeV}$ therefore
the peak position and height are not determined by the first resonance
only. 
The convergence for $E_p$ and $H_p$ in the consecutive orders 
of perturbation theory near the Coulomb solution 
is not fast in the $\MSsch$-scheme. 
For the typical numerical values of the theoretical parameters 
$m_t$, $\G_t$ and $\al_s(M_Z)$ one finds \cite{direg}
\ba
\label{conver}
E_p&=&E_0(1+0.58+0.38+\ldots)\nn \\
H_p&=&H_0(1-0.15+0.12+\ldots)
\ea 
(see also \cite{PY,Y1}).
Important contributions that affect the quality of
convergence are the local term ($\sim \al_s V_0\delta(\vec{r})$ 
which is related to $1/r^2$ non-Abelian term \cite{Gup})
and higher order PT corrections to $V_{pot}(r)$.

\section{Discussion and conclusion}
Slow convergence for the peak characteristics of the cross section
given in eq.~(\ref{conver}) has been actively discussed.
The suggestions of the redefinition of the top quark mass
have been made (e.g.~\cite{revHoa,recren,YakM})
as the use of the pole mass for a description of the cross section 
near the threshold is criticized 
on the ground of its infrared instability~\cite{mpoleIR}.
This approach is based on introduction of an effective mass 
that partly accounts for interaction~[41-43].
In this talk I only discuss some possible ways of
optimizing the convergence for the Green's function
with the pole mass as a theoretical parameter~\cite{talk1}.
Actual calculations near the threshold have been performed
within the pole mass scheme. For optimizing the convergence
one can use methods of exact summation of some contributions
in all orders and renormalization scheme invariance 
of PT series e.g.~\cite{KPP}.
The Hamiltonian can be written in the form 
\be
H=H_{LO}+\Delta V_{PT}+ \al_s V_0\delta(\vec{r}), 
\quad V_0=-\frac{70\pi}{9m_t^2}
\ee
where corrections are given by 
the perturbation theory corrections to $V_{pot}(r)$ 
($\Delta V_{PT}$-part) and by the local term
($\al_s V_0\delta(\vec{r})$-part).
The $\delta(\vec{r})$-part is a separable potential and can be taken 
into account exactly~\cite{nonrelPiv}.
The solution reads
\be
G(E;0,0)=\frac{G_{ir}(E;0,0)}{1+\al_s V_0 G_{ir}(E;0,0)}
\ee
with 
\be
G_{ir}(E)=(H_{LO}+\Delta V_{PT}-E)^{-1} 
\ee
being the irreducible Green's function.
The PT expansion of the static potential 
in NRQCD is important for getting
stable results for the cross section near the threshold 
because the static potential is the genuine quantity which is
computed in high order of PT in the strong coupling 
constant~\cite{av}. The convergence in the $\MSsch$ scheme 
is not fast which reflects the
physical situation that the observables 
represented by the cross section curve (for instance,
$E_p$ and $H_p$) are sensitive to
different scales. The finite-order perturbation 
theory expansion of the static potential 
cannot handle several distinct scales with the
same accuracy. Indeed, the PT expansion of the static potential 
is done near some (arbitrary) scale (or distance) 
which can be considered simply as a normalization point.
The farther a given point lies from this normalization point
the worse the precision of the PT expansion 
for the static potential is.
The PT expressions in the $\MSsch$ scheme 
are not directly sensitive to physical scales 
because subtractions are made in a mass independent way 
(for instance, massive particles do not decouple
automatically in the $\MSsch$ scheme e.g.~\cite{decoupl0,decoup}). 
One can rewrite the static potential through some physical 
parameters which is similar to the use of the momentum subtraction
scheme instead of the $\MSsch$-scheme 
\be
V_{pot}(r)=
-\frac{4\al_0}{3 r}\left(1+\al_0 b_1\ln r/r_0
+\al_0^2 (b_1^2\ln^2 r/r_0 +b_2'\ln r/r_0+c)\right)  \, .
\ee
Here $r_0$ and $c$ are the parameters of the renormalization scheme 
freedom in NNLO and $\al_0$ is the corresponding coupling in the
$\{r_0,c\}$-scheme~\cite{stevenson}. They parameterize 
the center of the expansion (a normalization point)
and the derivative (respective $\beta$-function)
of the static potential. The parameters ($r_0,c$) can be chosen such 
in order to minimize the higher order corrections to a particular 
observable (e.g.~\cite{KPP}
where NLO analysis has been done).
In such a case $r_0$ can be understood 
as a typical distance to which a chosen observable is sensitive. 
Note that the best approximation of the static potential $V_{pot}(r)$
for different scales would be provided by the use of 
the running coupling constant $\al_s(r)$.
The analytical calculation of GF
becomes technically impossible in this case. 
The numerical calculation of GF
requires some regularization at large distances
where the IR singularity (Landau pole) can occur in $\al_s(r)$.
The singularity can be dealt with if an IR fixed point appears 
in the evolution
for the effective coupling constant (e.g.~\cite{renorm})
or with some nonPT regularization for the potential
(e.g.~\cite{richard}).
For the top quark production the contribution of the 
large $r$ region into the cross section is small because of the large 
decay width of the top quark.
In the finite-order PT analysis the parameters 
$r_0$ and $c$ can be chosen to minimize higher order corrections
either to $E_p$ or to $H_p$ but not to both simultaneously
because $E_p$ and $H_p$ are sensitive to different distances.
One finds the difference of scales minimizing corrections 
to the first Coulomb resonance in NLO to be
\be
\label{relsc}
\ln(r_E/r_\psi)=\frac{1}{3}+\frac{\pi^2}{9} \, .
\ee
Because of the large top quark width many states contribute
into the position and height of the peak in the cross section.
Therefore the characteristic distance estimates are 
not so transparent (the NNLO peak position, for instance, is not exactly
the ground state energy in the zero width limit). 
The relation (\ref{relsc}) can serve just as a basic guide.
In practical analysis one can choose the particular numerical values for 
the parameters ($r_0,c$) which stabilize either $E_p$ or $H_p$. 

\section*{Acknowledgments}
I thank J.G.K\"orner for kind hospitality at Mainz University 
where the paper was completed.
The work is supported in part by Russian Fund for Basic Research 
under contracts 99-01-00091 and 01-02-16171, by INTAS and DFG grants.

\end{document}